\documentclass[prl,aps,twocolumn,superscriptaddress,lengthcheck,showpacs,amsmath,amssymb,floatfix,longbibliography]{revtex4-1}
\usepackage{graphicx}  
\usepackage{hyperref}
\usepackage{microtype}

\hypersetup{colorlinks,linkcolor=blue,citecolor=blue,urlcolor=blue}
\def\equationautorefname~#1\null{Eq.~(#1)\null}

\newcommand{\braket}[1]{\ensuremath{\left\langle{#1}\right\rangle}}
\newcommand{\bra}[1]{\langle #1 |}
\newcommand{\ket}[1]{| #1 \rangle}

\begin{document}

\title{Harvesting Excitons through Plasmonic Strong Coupling}
\author{Carlos Gonzalez-Ballestero}
\affiliation{Departamento de F{\'\i}sica Te{\'o}rica de la Materia Condensada and Condensed Matter Physics Center (IFIMAC), Universidad Aut\'onoma de Madrid, E-28049 Madrid, Spain}

\author{Johannes Feist}
\affiliation{Departamento de F{\'\i}sica Te{\'o}rica de la Materia Condensada and Condensed Matter Physics Center (IFIMAC), Universidad Aut\'onoma de Madrid, E-28049 Madrid, Spain}

\author{Esteban Moreno}
\affiliation{Departamento de F{\'\i}sica Te{\'o}rica de la Materia Condensada and Condensed Matter Physics Center (IFIMAC), Universidad Aut\'onoma de Madrid, E-28049 Madrid, Spain}

\author{Francisco J. Garcia-Vidal}
\affiliation{Departamento de F{\'\i}sica Te{\'o}rica de la Materia Condensada and Condensed Matter Physics Center (IFIMAC), Universidad Aut\'onoma de Madrid, E-28049 Madrid, Spain}
\affiliation{Donostia International Physics Center (DIPC), E-20018 Donostia/San Sebastian, Spain}


\begin{abstract}
Exciton harvesting is demonstrated in an ensemble of quantum emitters coupled to localized surface plasmons. When the interaction between emitters and the dipole mode of a metallic nanosphere reaches the strong coupling regime, the exciton conductance is greatly increased. The spatial map of the conductance matches the plasmon field intensity profile, which indicates that transport properties can be tuned by adequately tailoring the field of the plasmonic resonance. Under strong coupling, we find that pure dephasing can have detrimental or beneficial effects on the conductance, depending on the effective number of participating emitters. Finally, we show that the exciton transport in the strong coupling regime occurs on an ultrafast timescale given by the inverse Rabi splitting ($\sim10~$fs), orders of magnitude faster than transport through direct hopping between the emitters.
\end{abstract}

\pacs{71.35.-y, 05.60.Gg, 73.20.Mf, 81.05.Fb}
\maketitle


Bound electron-hole pairs in semiconductors and molecular solids, known as excitons, play a key role in many basic processes such as F\"orster resonance energy transfer or energy conversion in light-harvesting complexes~\cite{EngelNAT2007,CarusoChin2009,ScholesNATCHEM2011}. Various optoelectronic devices are also based on exciton dynamics, including organic solar cells~\cite{MenkeNATMAT2013}, light-emitting diodes~\cite{HofmannPRB2012}, and excitonic transistors~\cite{NovitskayaSCIENCE2008}. Thus, a large research effort is directed towards controlling exciton transport properties. As excitons usually suffer from relatively large propagation losses associated with decoherence and recombination, increasing their propagation length is an important goal.  Moreover, the transport of these quasiparticles into a specific spatial region in a controlled way could significantly increase the efficiency of devices such as solar cells, where the presence of excitons close to the charge separation region is essential for photocurrent generation.

Recent works~\cite{Johannes2014,Pupillo2014} have shown that exciton conductance in an ensemble of organic molecules can be boosted by several orders of magnitude when the ensemble is coupled to a cavity mode and the system enters the strong coupling (SC) regime. This is a promising result, which motivates the exploration of this phenomenon beyond cavity setups. Among the fields in which the SC regime is very attractive, plasmonics stands out due to the recent works studying SC between surface plasmons and quantum emitters (QEs)~\cite{BellessaPRL2004,SugawaraPRL2006,HalasNANOLETT2008,TruglerPRB2008,SridharanPRA2010,EbbesenPRL2011,VanVlackPRB2012,AlexPaloma2013,HalasNANOLETT2013,AlexPRL2014,TormaBarnes2015,Zengin2015}. Surface plasmons arise as ideal platforms for SC applications, due to their small mode volume and the tunability of their electric field profile via nanostructure design. These characteristics could allow for a deterministic control of exciton harvesting. 

In this Letter we demonstrate efficient harvesting of excitons in a collection of QEs strongly coupled to localized surface plasmons (LSPs). As a proof of principle, we first study a system of QEs interacting with the dipolar modes supported by a metal nanosphere (NS). We show how within the SC regime the spatial map of the exciton conductance mirrors the field intensity profile. Employing a structure composed of three aligned nanospheres, we show that the efficiency of exciton harvesting can be significanly increased by tuning the electric field profile of the LSP mode. In addition, we demonstrate that the role of dephasing in the exciton conductance strongly depends on the effective number of emitters involved in the formation of the polariton modes. Finally, we demonstrate that the speed of exciton transport in the SC regime is orders of magnitude faster than in the weak coupling regime. 

\begin{figure}
\centering
\includegraphics[width=\linewidth]{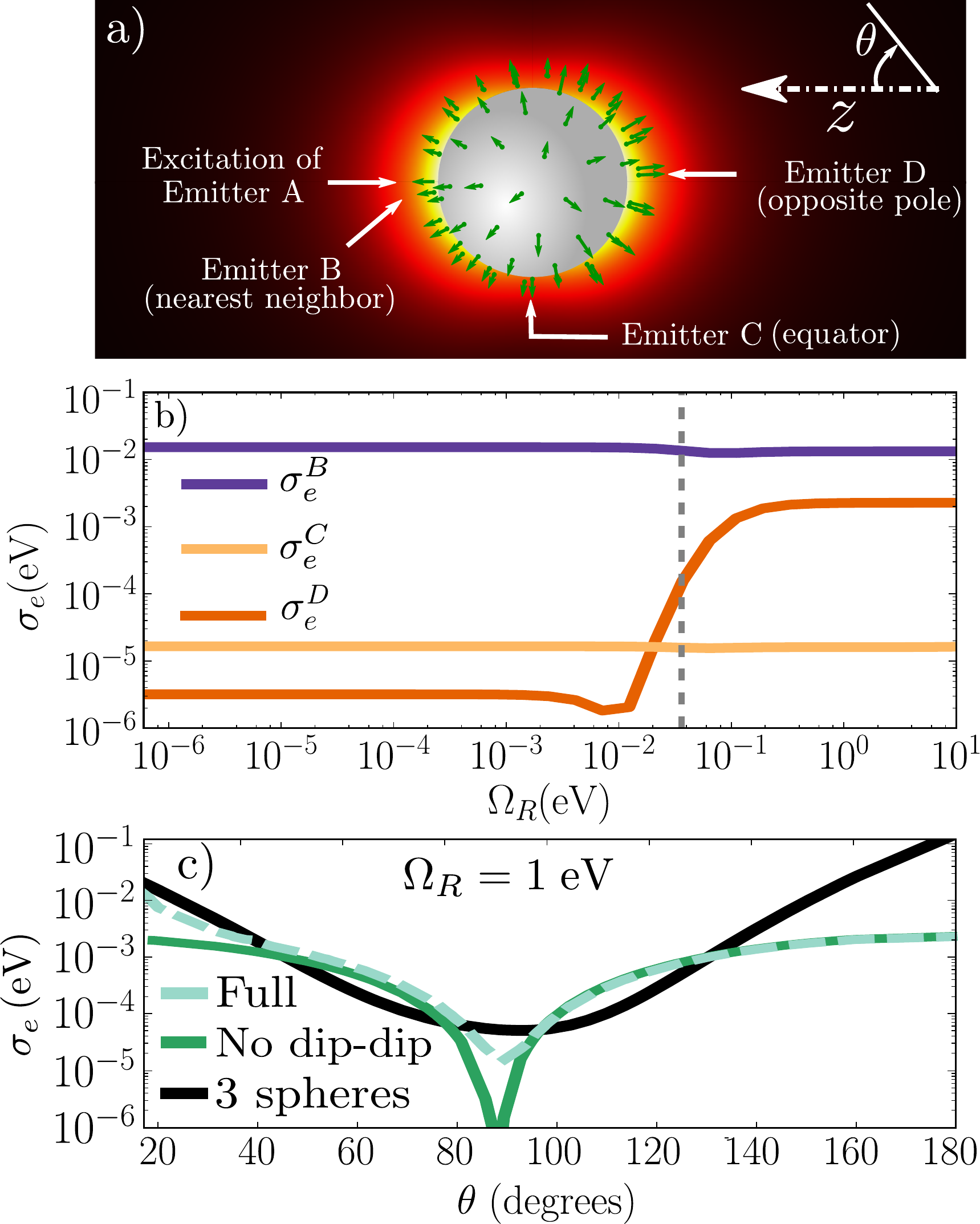}
\caption{(color online) a) Illustration of the single-NS system. The colored background shows the electric field intensity associated with the $z$-oriented dipole mode. b) Exciton conductance versus Rabi splitting for $N=100$ emitters. The gray dashed line indicates the onset of SC, $\Omega_R \gtrsim \vert \gamma - \kappa \vert/2$. c) Angular dependence of the conductance in the SC regime (light green) and the same quantity when neglecting dipole-dipole interaction between QEs (dark green). For comparison the three-NS case is also depicted (black line).}\label{figg1}
\end{figure}

The first system we consider consists of a silver NS of radius $R$, surrounded by a layer of $N$ QEs, as shown in \autoref{figg1}a. For simplicity, the QEs are regularly distributed over a spherical layer, which we place at a distance $h=1~$nm away from the surface of the NS. Note that at this short distance, higher multipole modes of the NS are dominant and lead to quenching losses for a single QE~\cite{anger2006}. However, recent works have shown that for $N$ emitters, collective SC with the dipole resonance of the NS does indeed arise, and higher multipoles merely add an effective detuning to the hybrid mode~\cite{AlexPRL2014}. Hence, for the silver nanosphere we only consider the three dipolar LSP modes ($x,y,z$), which are characterized by their frequency $\omega_{pl}$ and decay rate $\kappa$, and electric field profile $\vec{E}_\alpha(\vec{r})$ ($\alpha\in x,y,z$). These parameters can be extracted from the NS properties, as described in the Supplemental Material~\cite{suppl}. The QEs are modelled as point dipoles oriented along the radial direction, with transition frequency $\omega_0$, dipole moment $\vec{\mu}$, and total decoherence rate $\gamma = \gamma_\phi + \gamma_d$, where $\gamma_\phi$ accounts for pure dephasing, while the decay rate $\gamma_d = \gamma_r + \gamma_{nr}$ contains radiative and nonradiative contributions.

The Hamiltonian associated with the emitters-NS system within the rotating wave approximation can be written as
\begin{equation}\label{hamiltonian}
\begin{split}
H = & \sum_j \omega_0 c_j^\dagger c_j +\sum_{\alpha=x,y,z}\omega_{pl} a_\alpha^\dagger a_\alpha +\\
 &\sum_{i\ne j}V_{ij}(c_i^\dagger c_j + H.c.) + \sum_{j,\alpha }(g_{j\alpha} c_j^\dagger a_\alpha+ H.c.).
\end{split}
\end{equation}
Here, the operators $c_j$ and $a_\alpha$ annihilate an excitation in emitter $j\;(=1,...,N)$ and the LSP mode $\alpha$, respectively. The dipole-dipole interaction between QEs is given by $V_{ij}$, and the QE-LSP coupling is $g_{j\alpha} = -\vec{\mu}\cdot \vec{E}_\alpha(\vec{r}_j)$. As we will show later, in the SC regime only a single LSP mode contributes and dipole-dipole interactions can be neglected. Under these approximations and for zero detuning ($\omega_0=\omega_{pl}$), the $N+1$ singly excited eigenstates of $H$ are formed by: i) two polaritons $\ket{\pm}=\frac{1}{\sqrt{2}}(a^\dagger\ket{0}\pm\ket{B})$, where $\ket{B} = \frac{2}{\Omega_R}\sum_i g_i c_i^\dagger\ket{0}$ is the collective molecular bright state, with $\Omega_R$ the Rabi splitting ($\Omega_R^2 = 4\sum_j |g_j|^2$), and ii) the so-called dark states, $N-1$ combinations of molecular excitations orthogonal to $\ket{B}$ which have no LSP component.
The eigenfrequencies of the two polaritons are $\omega_0\pm \Omega_R/2$.

In order to study exciton transport through the ensemble of emitters, we first determine the steady state of the system when one of the QEs (emitter $A$ from now on) is incoherently pumped. Notice that this is the only driving term and no additional external illumination is present.
The system is described by its density operator $\rho$, whose dynamics is governed by the master equation $\dot{\rho}=-i[H,\rho]+\sum_j\gamma\mathcal{L}_{c_j}+\sum_\alpha\kappa\mathcal{L}_{a_\alpha}+\gamma_p\mathcal{L}_{c_A^\dagger}$. Incoherent processes (losses and pump) are described by Lindblad terms, $\mathcal{L}_b[\rho] = b\rho b^\dagger-\{b^\dagger b,\rho\}/2$. In the first part of this work we treat pure dephasing as a decay channel for simplicity. The pumping rate $\gamma_p$ is chosen small enough to stay in the linear regime.  The steady state density matrix $\rho_{ss}$ is obtained numerically using the open-source QuTiP package~\cite{JohanssonQUTIP2013}. Finally, the exciton conductance measuring the efficiency of exciton transfer from emitter $A$ to emitter $j$ is calculated as $\sigma_e^{(j)} = J_{(A\to j)}/\gamma_p$, where $J_{(A\to j)} = \gamma \operatorname{Tr} \left(H\mathcal{L}_{c_j}[\rho_{ss}]\right)$ is the energy loss rate of emitter $j$~\cite{Johannes2014}.

The exciton conductance is shown in \autoref{figg1}b for QEs at three representative positions, $B$, $C$, and $D$, as depicted in \autoref{figg1}a. With organic molecule applications in mind, the parameters of the $N=100$ QEs are chosen to correspond to TDBC J-aggregates at room temperature~\cite{MollJCHEMPHYS1995,ValleauJCHEMPHYS2012,SchwartzCPC2013}: $\omega_0 = 2.11~$eV, $\mu = 0.749~$e$\cdot$nm, $\gamma_\phi= 26.3~$meV, $\gamma_r = 1.32 \cdot 10^{-6}~$eV, $\gamma_{nr} = 1.10~$meV. The nanosphere (radius $R = 10~$nm) is embedded in a dielectric host of permittivity $\epsilon_d=6.8$, so that the LSPs are in resonance with the QEs. Finally, the plasmon losses are given by $\kappa = 0.1~$eV. In order to study different Rabi frequencies $\Omega_R$, we first artificially tune the field strength of the LSPs, instead of varying the number of QEs as could be done experimentally. Our results show how the onset of SC clearly differentiates two different regimes for exciton transport. In the weak coupling case (small $\Omega_R$), the dipole-dipole interaction between the QEs governs the dynamics and the transport is rather inefficient over large distances. The exciton conductance to emitter $D$ is thus smaller than that to $B$ or $C$. In the SC regime, however, this situation changes drastically, as LSP-mediated interaction becomes the primary transport channel~\cite{Johannes2014}. Due to the dipolar field profiles, emitter $A$ only couples to the $z-$dipole mode of the NS, and the $x$ and $y$ dipole play a negligible part. As the couplings $g_{j\alpha}$ are proportional to the electric field of mode $\alpha$, excitons are transferred more efficiently to regions of high field intensity. This is the cause of the boost in the conductance of emitter $D$ displayed in \autoref{figg1}b. 

These results indicate that the strong coupling conductance is position-dependent, mimicking the field intensity profile. This fact is confirmed in \autoref{figg1}c. Here, we plot the conductance in the SC regime for every QE versus its polar angular coordinate $\theta$. The results without dipole-dipole coupling ($V_{ij}=0$) agree very well with the full calculation. The significant dip of $\sigma_{e}$ around $\theta=90^{\circ}$ confirms that only the $z$-oriented LSP mode plays a relevant role. Therefore, we can safely neglect both the dipole-dipole interactions and the $x$- and $y$-oriented LSP modes within the SC regime. Under these approximations, an analytical solution for the master equation can be obtained~\cite{suppl}. For zero detuning ($\omega_{pl}=\omega_0$), the exciton conductance between emitters $A$ and $j$ has the simple form
\begin{equation}\label{sigmaNS}
\sigma_e^{(j)} = \frac{16\vert g_{A}\vert^2 \vert g_{j}\vert^2 \omega_0\left(\gamma + \Gamma\right)}{\Gamma\left(\Omega_R^2 + \gamma\kappa\right)\left(\Omega_R^2+2\gamma\Gamma\right)},
\end{equation}
where we have defined the rate $\Gamma = \gamma + \kappa$, and $g_j \equiv g_{jz}$ for simplicity.
Up to now, we have considered the case in which both the pumping and collection involve a single QE, and $\Omega_R$ is artificially modified by tuning the field strength of the LSP while keeping $N$ constant. In a realistic experiment, several emitters near location $A$ would be pumped, and excitons collected from a region around $D$. Furthermore, $\Omega_R\propto\sqrt{N}$ would be varied by changing the number of emitters $N$. \hyperref[sigmaNS]{Equation (\ref*{sigmaNS})} is easily generalized to this case, giving
\begin{equation}\label{sigmaNS_2}
\sigma_e = \frac{\eta_A \eta_D \omega_0\left(\gamma + \Gamma\right)\Omega_R^4}{\Gamma \left(\Omega_R^2 + \gamma\kappa\right) \left(\Omega_R^2+2\gamma\Gamma\right)},
\end{equation}
where $\eta_X = 4\sum_{j\in X} |g_j|^2 / \Omega_R^2$ measures the fraction of the Rabi frequency due to emitters involved in the pumping ($\eta_A$) and collection processes ($\eta_D$), respectively. Both $\eta_A$ and $\eta_D$ are independent of $N$ for uniform distributions of emitters. For small $N$ (i.e., weak coupling), \autoref{sigmaNS_2} shows that $\sigma_e$ grows as $N^2$, while when the SC regime is entered for large $N$ ($\Omega^{2}_{R} \gg \gamma\Gamma$), it saturates to a constant value, $\eta_A\eta_D \omega_0(1+\gamma/\Gamma)$. This equation thus predicts that the exciton transport efficiency from a pumping site to a collection spot can be increased by tailoring the mode to have maximal field strength at (only) these two locations in order to have large $\eta_A$ and $\eta_D$.

\begin{figure}
  \centering
  \includegraphics[width=\linewidth]{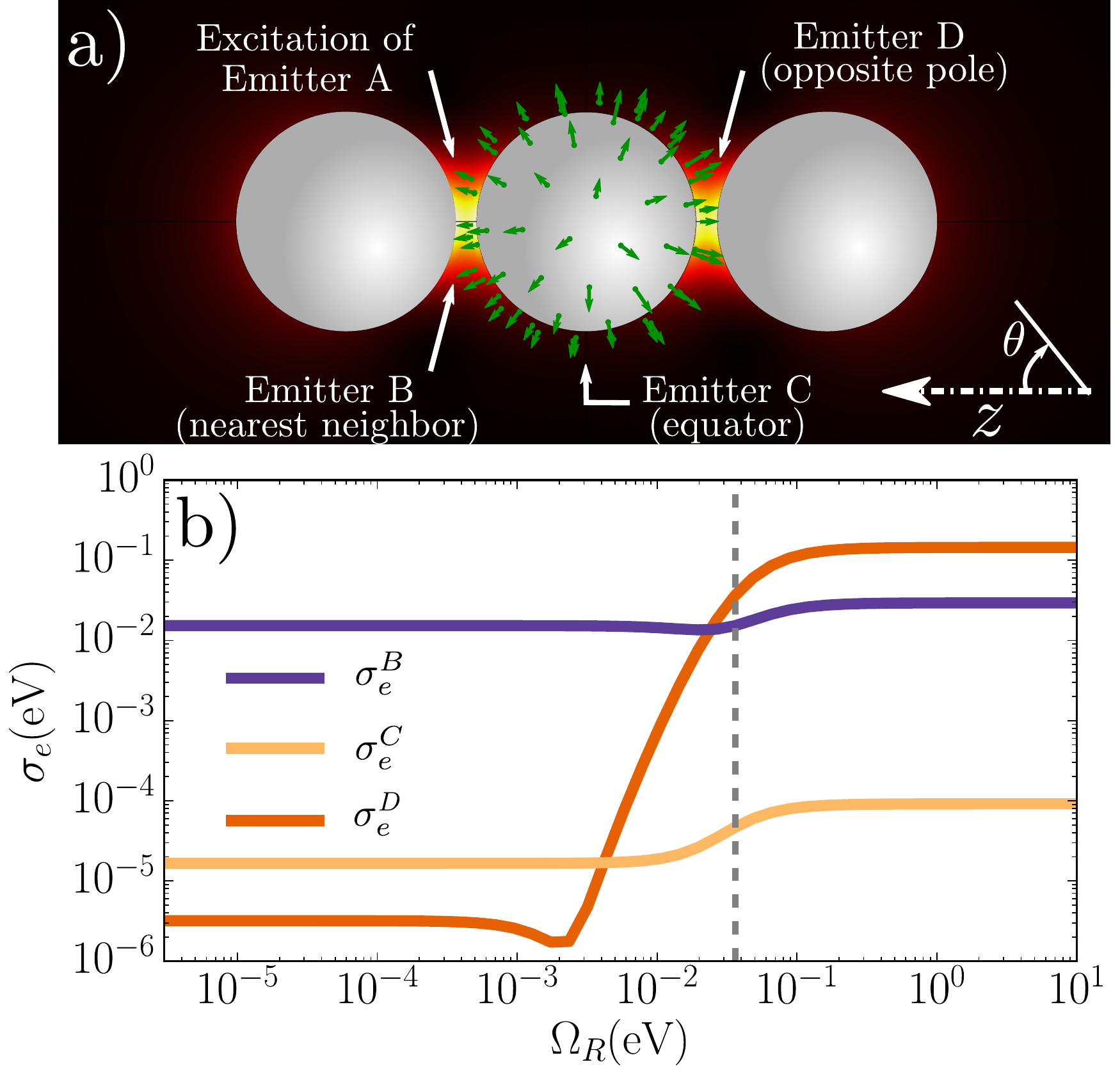}
  \caption{(color online) a) Illustration of the three-NS system. The colored background shows the electric field intensity of the lowest energy mode. b) Exciton conductance versus Rabi splitting for $N=100$ emitters. The curves correspond to transport from emitter $A$ to emitters $B$, $C$, and $D$ as depicted in panel (a). As in \autoref{figg1}b, the gray dashed line indicates the onset of SC.}\label{figg2}
\end{figure}

We demonstrate this by adding two additional identical silver nanospheres to the existing structure, as shown in \autoref{figg2}a. The background of the panel displays the field intensity map of the lowest energy mode, which in this case is not degenerate as the rotational symmetry is broken. The spheres are separated by a $2~$nm gap. In order to facilitate the comparison with the single-NS case and focus on the effect of the different mode profile, the LSP frequency and losses as well as the QE properties and locations are kept unchanged. Furthermore, we again plot the single-emitter to single-emitter conductance. The exciton conductance as a function of the Rabi splitting is displayed in \autoref{figg2}b for $N=100$. As the figure shows, the onset of SC again produces a substantial increase in the pole-to-pole conductance $\sigma_e^D$, significantly larger than in the single-NS case. The most striking feature of this system is that transport to emitter $D$ is now much more efficient than to the nearest neighbor $B$ of emitter $A$. The conductance in the SC regime is much more concentrated around the poles than in the single-NS (see \autoref{figg1}c), and conductance to point $D$ is increased by a factor of $50$. Excitons are thus shown to be very efficiently harvested at the \textit{hot spots} of the LSP mode.
This is an interesting result towards potential applications, due to the high tunability provided by the wide variety of plasmonic nanostructures that are available nowadays.

Up to now, our treatment of relaxation processes has been relatively crude, lumping together dissipation and dephasing. Recent works show that dephasing mechanisms can be relevant for exciton transport in organic compounds~\cite{CarusoChin2009}. Therefore, we now describe pure dephasing in more detail by performing the substitution $\gamma\mathcal{L}_{c_j} \to \gamma_d\mathcal{L}_{c_j}+\gamma_\phi\mathcal{L}_{c_j^\dagger c_j}$ in the master equation. In order to avoid unrealistic conclusions, we have also checked that the complementary Bloch-Redfield-Wangsness formalism~\cite{petruccione2002theory} reproduces the general behavior obtained with the standard Lindblad method. 

\begin{figure}
\centering
\includegraphics[width=\linewidth]{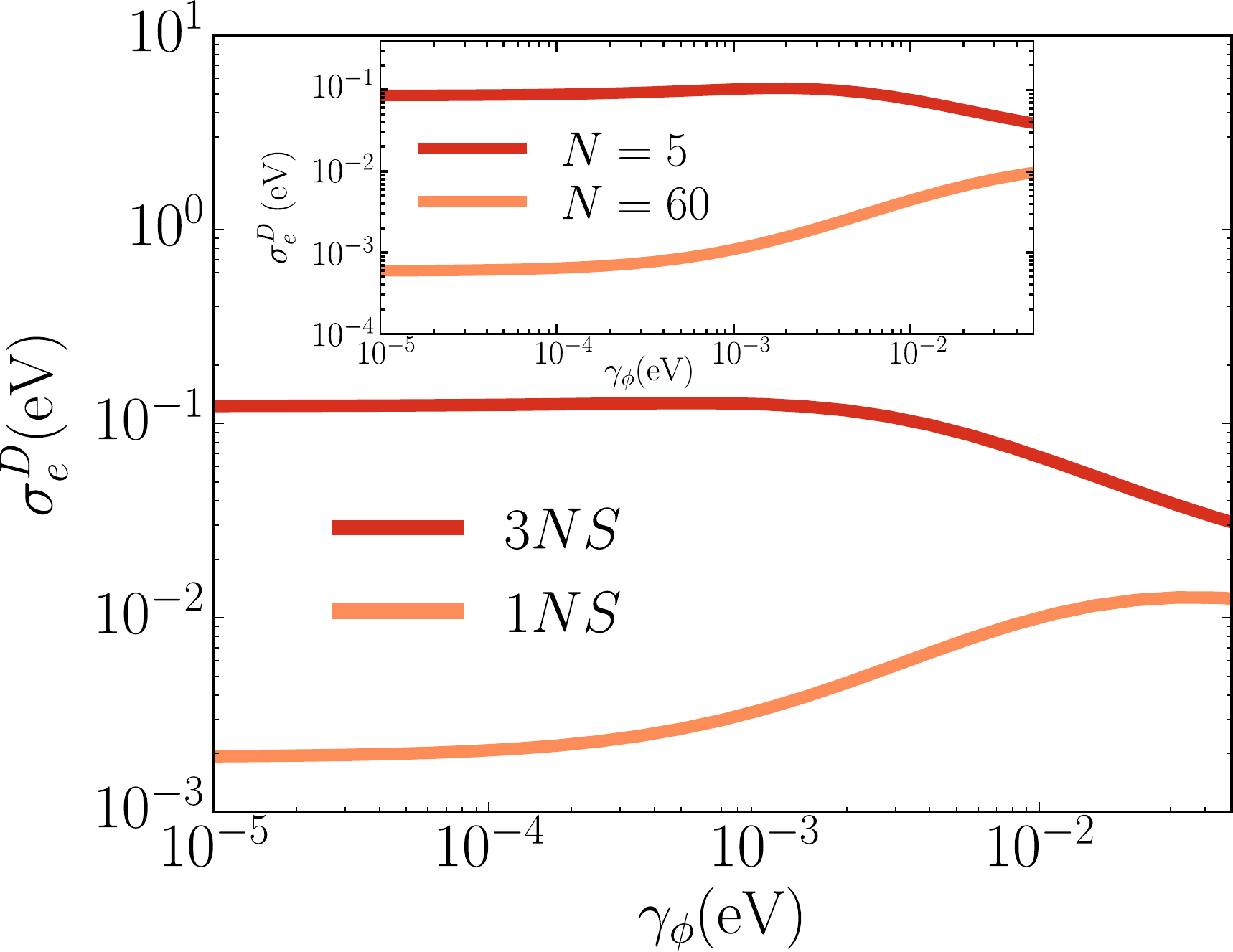}
\caption{Exciton conductance from point $A$ to point $D$ in the SC regime ($\Omega_R = 1~$eV), as a function of the dephasing rate. The single-NS case (orange line) is compared with the three-NS structure (red line). The inset displays the conductance in the SC regime for a system of QEs uniformly coupled to the field, in the cases $N=5$ and $N=60$.}\label{figg3}
\end{figure}

The pole-to-pole exciton conductance in the SC regime is shown in \autoref{figg3} as a function of the dephasing rate, for both the single-NS and the three-NS cases. Except for the value of $\gamma_\phi$, the parameters of the three nanospheres and the QEs are the same as in the previous calculations. Surprisingly, the dependence of the conductance with dephasing is remarkably different for the two considered nanostructures. As dephasing is increased, the conductance decreases monotonically for the three-NS structure, but counterintuitively increases for the single NS. As we show next, this difference in behavior is related to the effective number of QEs that enter strong coupling with the LSP. This number is quite small in the three-NS case, where only QEs near the hot spots participate in SC. Our hypothesis is confirmed by considering a system of $N$ non-interacting QEs uniformly coupled to a cavity mode ($g_j = g$). In this simple situation, an analytical formula for the exciton conductance can be derived~\cite{Johannes2014}, which in the SC limit is given by
\begin{equation}\label{sigmaHOMO}
\sigma_{e}=\frac{\omega_0\gamma_d(\gamma_d+\gamma_\phi)(\kappa+2\gamma_d+2\gamma_\phi)}{\left(2\gamma_\phi + \gamma_d N\right)\left(\kappa\gamma_\phi + \gamma_d N (\kappa + \gamma_d + \gamma_\phi)\right)}.
\end{equation}

This simple expression (shown in the inset of \autoref{figg3}) is able to reproduce the observed features. Specifically, the ratio $\sigma_{e}(\gamma_\phi\!\to\!\infty)/\sigma_{e}(\gamma_\phi\!\to\! 0)$ is approximately equal to $\gamma_dN^2/\kappa$. Since typical plasmonic structures fulfill $\gamma_d/\kappa \ll 1$, this expression predicts that with increasing dephasing, exciton conductance decreases for small $N$ ($N<\sqrt{\kappa/\gamma_d}$), but increases for sufficiently large $N$ ($N>\sqrt{\kappa/\gamma_d}$). This dependence with the number of emitters suggests that the dark states play a key role in this process. Since dephasing creates an incoherent coupling between the dark modes and the polaritons, a fraction of the population in the dark modes can be transferred to the polaritons. For large $N$, the dark states are highly populated as the overlap between the initial state (one excitation at emitter A) and the polaritons is extremely small. As a consequence, the excitation transfer to the polaritons can compensate for the detrimental effect of dephasing on this state, making the conductance increase.

\begin{figure}
\centering
\includegraphics[width=\linewidth]{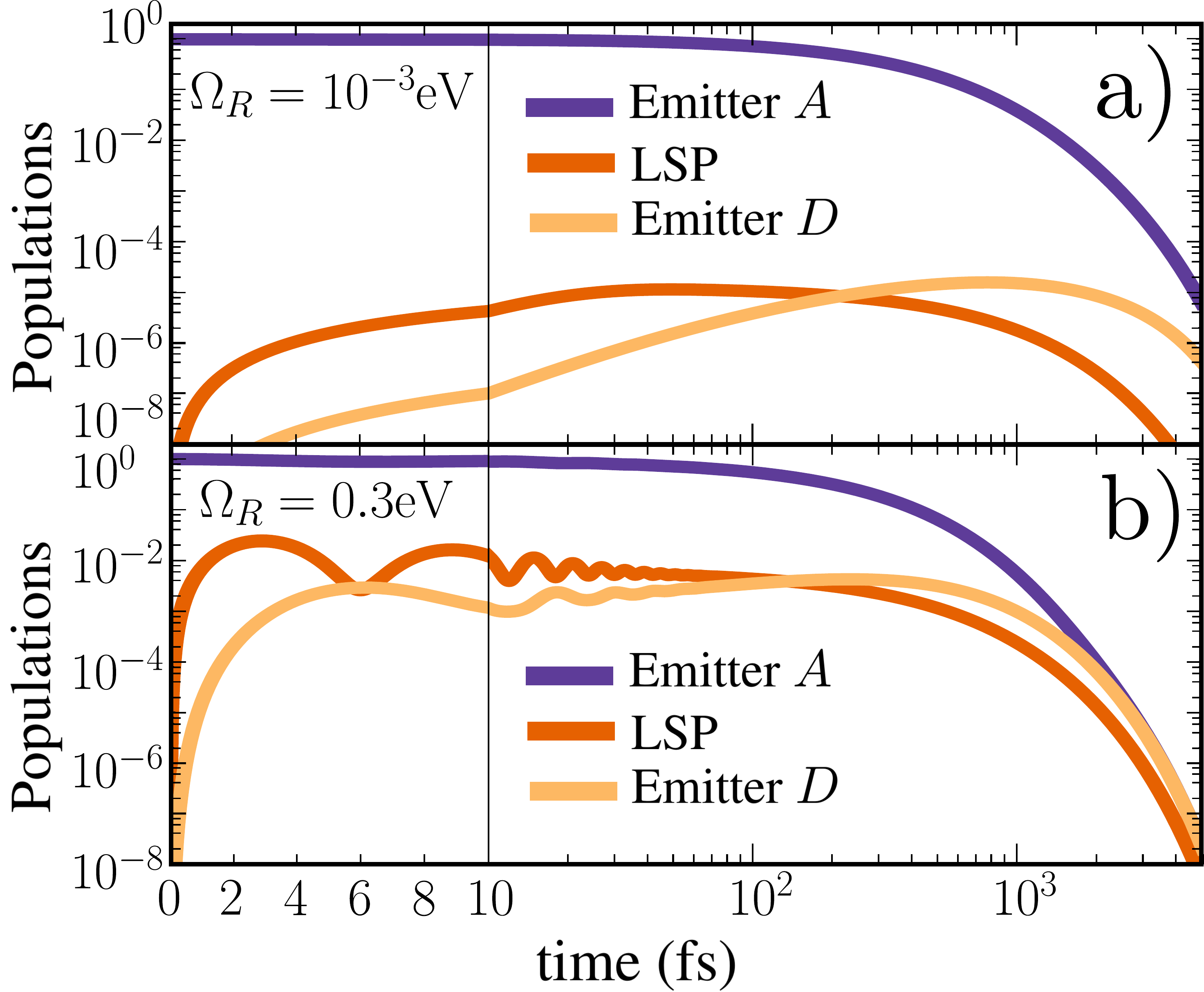}
\caption{(color online). Time dynamics of the single NS system when QE $A$ is initially excited (initial state $c^\dagger_A\ket{0}$) and no pump is applied. The populations of both the main LSP and the relevant emitters, $A$ and $D$, are displayed as a function of time. a) Weak coupling case, $\Omega_R = 10^{-3}~$eV. b) Strong coupling case, $\Omega_R = 0.3~$eV. In both panels, the linear scale for short times allows for a better visualization of the results.}
\label{figg4}
\end{figure}

While we have so far focused on the conductance as obtained in the steady state under pumping, the temporal dynamics of the system provides important additional insight. We thus investigate the population dynamics in the single NS case, for an initial excitation of emitter A. In the weak coupling regime, the dipole-dipole interaction dominates and slowly transfers population to emitter D (as shown in \autoref{figg4}a). The plasmon modes do not significantly participate in the dynamics, so that all populations decay with the lifetime of the bare QEs ($\tau \sim 600~$fs) for large times. As the exciton transport is even slower, the increase of population in emitter D is cut off at around this time, with a maximum population of $\sim2\cdot10^{-5}$. On the other hand, in the SC regime (\autoref{figg4}b), the population is delivered to the emitter D much more efficiently through Rabi oscillations. These proceed on a timescale determined by the inverse of the Rabi splitting ($1/\Omega_R \sim 15~$fs in \autoref{figg4}b), giving extremely fast population transfer. Furthermore, the population of emitter D reaches significantly larger values than in the weak coupling regime, up to $\sim4\cdot10^{-3}$. For large times, most of the population is trapped in the dark states, which can now also decay through the dephasing-induced coupling to the polaritons, giving an effective lifetime somewhat below the bare QEs.

In conclusion, we have demonstrated the feasibility of harvesting excitons through strong coupling in systems composed of a plasmonic structure and an ensemble of organic molecules. We have shown that for emitters coupled to the localized surface plasmons of a single nanoparticle, the exciton conductance map mimics the electric field profile of the plasmon resonance. Taking advantage of this property, we have devised a more complex structure in which an exciton can be efficiently transferred between two subwavelength hot-spots of the plasmonic system. We have also shown how dephasing can be beneficial or detrimental depending on the number of emitters that are effectively coupled to the plasmon resonance. We have additionally demonstrated that exciton transport in the strong coupling regime proceeds orders of magnitude faster than under weak coupling. Finally, it is worth noting that our findings regarding harvesting of excitons mediated by strong coupling are general and applicable to any quantum emitter, from atoms and quantum dots to organic molecules, and also to any confined electromagnetic mode with similar properties as the plasmonic ones used here.

\begin{acknowledgments}
This work has been funded by the European Research Council (ERC-2011-AdG Proposal No. 290981), the Spanish MINECO (FPU13/01225 fellowship and MAT2011-28581-C02-01 grant), and by the European Union Seventh Framework Programme under grant agreement FP7-PEOPLE-2013-CIG-618229.
\end{acknowledgments}

\bibliography{bibliography}

\pagebreak

\onecolumngrid
\begin{center}
\textbf{\large Supplemental Material}
\end{center}
\setcounter{equation}{0}
\setcounter{figure}{0}
\setcounter{table}{0}
\setcounter{page}{1}
\makeatletter

\vspace{0.9cm}

\twocolumngrid

\section{Calculation of the mode properties}
We calculate the classical field profiles and modal characteristics of the localized surface plasmon (LSP) numerically with the finite element method (using COMSOL Multiphysics) to solve Maxwell's equations. The permittivity of the silver structures is given by a Drude-Lorentz formula:
\begin{equation}\label{Drude}
\epsilon(\omega) = \epsilon_\infty -\frac{\omega_p^2}{\omega(\omega + i\gamma_D)}-\Delta\frac{\Omega_P^2}{\omega^2 - \Omega_P^2 +i\omega\Gamma_P},
\end{equation}
where the parameters $\epsilon_\infty\!=\!3.91$, $\omega_p\!=\!8.833\ $eV, $\gamma_D\!=\!0.0553\ $eV, $\Delta\!=\!0.76$, $\Omega_P\!=\!4.522\ $eV, and $\Gamma_P\!=\!8.12\ $eV are taken from Ref.~\cite{zhiming2008}. For the dipole mode of the single NS, we obtain a frequency $\omega_{pl} = 2.11\ $eV and a dissipation rate $\kappa = 0.12\ $eV. In the three-NS structure, the parameters of the lowest energy mode are $\omega_{pl}' = 1.85 \; \mathrm{eV}$, and $\kappa' = 0.14 \; \mathrm{eV}$. Note that in the main text these values are modified artificially to agree with the single-NS case, in order to isolate the influence of the electric field profile.

\section{Quantization of LSP fields}

The classical fields of the system modes obtained from Maxwell's equations have to be properly quantized in order to include them in a quantum Hamiltonian. In this section we briefly detail the method followed for this purpose.

\subsection{Single-sphere case}

The dipole mode of a single sphere is quantized by comparing the classical and quantum values of the polarizability $\alpha$. For a small metallic sphere of permittivity $\epsilon(\omega)$ and radius $R$, the classical static polarizability is given by \cite{bohren2008a}
\begin{equation}\label{alphaCL}
\alpha_{cl} = 4\pi \epsilon_0 \epsilon_d R^3 \frac{\epsilon(\omega)-\epsilon_d}{\epsilon(\omega)+2\epsilon_d},
\end{equation}
being $\epsilon_d$ the permittivity of the surrounding dielectric. For simplicity, we introduce an approximate Drude permittivity (\autoref{Drude}, with $\Delta = 0$). We can define  a resonance frequency $\omega_r = \omega_p/\sqrt{\epsilon_\infty + 2\epsilon_d}$ and, after some algebra \cite{SridharanPRA2010}, arrive to the following expression:
\begin{equation}\label{alphaCL2}
\alpha_{cl} \approx -2\pi \epsilon_0 \epsilon_d R^3 \frac{3\epsilon_d}{\epsilon_\infty + 2\epsilon_d} \frac{\omega_r^2}{\omega(\omega-\omega_r+i\gamma_D/2)},
\end{equation}
which is valid in the vicinity of a narrow resonance, i.e. $\omega \approx \omega_r \ll \gamma_D$.

On the other hand, the polarizability of a quantum two-level system with dipole moment $\mu$ is given by \cite{loudon2000quantum}:
\begin{equation}
\alpha_{q} = \frac{\mu^2}{\hbar}\frac{2\omega_0}{\omega_0^2 - \left(\omega+i\gamma_0/2\right)^2},
\end{equation}
where $\omega_0$ and $\gamma$ are the two-level system frequency and linewidth, respectively. Close to a narrow resonance ($\omega \approx \omega_0 \ll \gamma_0$), the above expression is approximated as 
\begin{equation}\label{alphaTLS}
\alpha_{q}  \approx -\frac{\mu^2}{\hbar}\frac{\omega_0}{\omega\left(\omega - \omega_0+i\gamma_0/2\right)}.
\end{equation}
By direct comparison of Eqs.~(\ref{alphaCL2}) and (\ref{alphaTLS}), we obtain the dipole moment of a nanosphere in a quantum model:
\begin{equation}
\mu_{pl} = \left(\hbar \omega_{pl} \frac{6\pi\epsilon_0\epsilon_d^2R^3}{\epsilon_\infty + 2\epsilon_d}\right)^{1/2}. 
\end{equation}
The quantum electric field in this case corresponds to the classical field emitted by a dipole with dipole moment $\mu_{pl}$.

\subsection{Arbitrary geometry}

For more complex nanostructures, the eigenmodes do not possess a purely dipolar profile, and a new quantization procedure is necessary. We start with the classical electric and magnetic field profiles obtained from our simulations, $\vec{E}_{cl}(\vec{r})$ and $\vec{H}_{cl}(\vec{r})$. We generalize the usual quantization procedure for vacuum fields \cite{loudon2000quantum}, by applying the correspondence principle to the electromagnetic energy:
\begin{equation}\label{correspondence}
U\left(\vec{E}_{cl},\vec{H}_{cl}\right) \leftrightarrow H\left(\vec{E}_{q},\vec{H}_{q}\right).
\end{equation}
In the above expression, $U$ stands for the total classical electromagnetic energy, and $H$ for the Hamiltonian operator. The quantum field operators $\vec{E}_{q},\vec{H}_{q}$ are related to their classical counterparts,
\begin{equation}\label{Qfields}
\begin{split}
\vec{E}_{q} & = C\vec{E}_{cl} a+ C^*\vec{E}_{cl}^* a^\dagger,\\
\vec{H}_{q} & = C\vec{H}_{cl} a+ C^*\vec{H}_{cl}^* a^\dagger,
\end{split}
\end{equation}
where $a$ and $a^\dagger$ are the photonic mode annihilation and creation operators, respectively, and $C$ is a normalization constant we have to determine.

Our systems are composed of a metallic nanostructure surrounded by a dielectric host, with permittivities $\epsilon(\omega)$ and $\epsilon_d$ respectively. Usually, the electromagnetic energy $U$ is ill-defined in lossy media, and a macroscopic QED formalism is required. However, in the vicinity of a narrow resonance $\omega = \omega_{pl}'$, and provided that the losses are small $\left(\mathrm{Im}[\epsilon(\omega)] \ll \mathrm{Re}[\epsilon(\omega)] \right)$, the following approximation holds \cite{maier2007plasmonics}\\
\vspace{-0.4cm}
\begin{equation}\label{Ucl}
\begin{split}
U & \approx \frac{\epsilon_0}{2} \int_{V_{\mathrm{metal}}} dV  \mathrm{Re}\left[\frac{d(\omega\epsilon(\omega))}{d\omega}\right]_{\omega = \omega_{pl}'} \vec{E}_{cl}^* \cdot \vec{E}_{cl} \\
+& \frac{\epsilon_0\epsilon_d}{2} \int_{V_{\mathrm{diel}}} dV  \vec{E}_{cl}^* \cdot \vec{E}_{cl}+\frac{\mu_0}{2} \int_{V_{\mathrm{tot}}} dV  \vec{H}_{cl}^* \cdot \vec{H}_{cl}.
\end{split}
\end{equation}
We apply the correspondence principle, \autoref{correspondence}, by introducing the quantum fields (\ref{Qfields}) in the above equation, i.e. substituting $(\vec{E}_{cl}^* , \vec{E}_{cl},\vec{H}_{cl}^* , \vec{H}_{cl})$ by $(\vec{E}_{q}^\dagger , \vec{E}_{q},\vec{H}_{q}^\dagger , \vec{H}_{q})$. After manipulation, we arrive to the familiar Hamiltonian
\begin{equation}\label{H}
H = \vert C \vert^2 U \left(2a^\dagger a + 1\right). 
\end{equation}
The comparison with the Hamiltonian of a harmonic oscillator, $H = \hbar\omega_{pl}'(a^\dagger a + 1/2)$, gives the expression of the normalization constant,
\begin{equation}
C =  \sqrt{\frac{\hbar\omega_{pl}'}{2U}}.
\end{equation}
Finally, after $C$ is determined, the calculation of the couplings $g_j' = -\vec{\mu}\cdot C\cdot\vec{E}_{cl}(\vec{r}_j)$ is straightforward.

It is important to mention that, when calculating the energy $U$ from our simulations, we have to deal with the linear divergence of the second integral in \autoref{Ucl}. This is a fundamental problem regarding lossy cavities, in which normal modes are ill-defined. In our case, we can safely ignore the divergent contribution due to the low loss rate~\cite{KoenderinkOptLett2010}. The validity of this approximation has been checked in the single nanosphere case, where this method reproduces the analytical results obtained above very accurately.

\section{Analytical formula for the conductance}

Here we give details of the calculation of the formula for the conductance (Eq.~(2) in the main text). In the absence of dipole-dipole interaction, and considering only one of the dipole modes, the Hamiltonian of the system is given by the Tavis-Cummings expression,
\begin{equation}\label{Hamiltonian}
H = \omega_0\sum_{j=1}^{N} c_j^\dagger c_j +  \omega_{pl} a^\dagger a + \sum_{j}\left(g_{j} a^\dagger c_j + \mathrm{H.c.}\right).
\end{equation}
It is useful to work in the bright-dark basis, which is composed of a bright state $\ket{B}$,
\begin{equation}\label{bright}
\ket{B} = \frac{2}{\Omega_R}\sum_{j=1}^{N}g_jc_j^\dagger \ket{0},
\end{equation}
and a set of $(N-1)$ dark states. In principle, we are free to choose any orthonormal set of states, and we use this freedom to our advantage. As in our problem we pump the first molecule (state $c_1^\dagger\ket{0}$), we choose the dark states such that this pumping only excites one of them, which we name $\ket{D}$. It is straightforward to prove that
\begin{equation}\label{dark}
  \ket{D} = \frac{2g_1}{G'\Omega_R}\left(\frac{G'^2}{g_1^*}c_1^\dagger - \sum_{j=2}^Ng_jc_j^\dagger\right)\ket{0},
\end{equation}
 where we have defined $G'^2 = (\Omega_R/2)^2 - g_1^2$ for simplicity. The remaining dark states $\lbrace  \ket{D_k}  ; \; k=1,...N-2\rbrace$ are arbitrary, provided that they fulfill the orthogonality conditions $ \braket{D_k|B} = \braket{D_k|D} = 0$. Note that Eqs.~(\ref{bright}) and (\ref{dark}) imply that
$\bra{D_k} \sigma_1^\dagger \ket{0} = \bra{D_k} (a\ket{B} + b\ket{D}) =0$.

Consequently, the states $\bra{D_k}$ are not coupled to the pumped state and do not take part in the dynamics. The time evolution of the system is thus restricted to the 4-dimensional subspace spanned by the states $\lbrace \ket{0}, \ket{B}, \ket{D}, a^\dagger \ket{0}\rbrace$, and the master equation reduces to a $16\times 16$ linear system. 

We can thus calculate the steady-state density matrix analytically, from which it is straightforward to obtain the exciton conductance. With the notation employed in the main text, the general expression is given by
\begin{widetext}
\begin{equation}
\sigma_e(j) = -16\vert  g_j\vert^2\vert g_A\vert^2 \frac{\gamma \delta (\kappa \Omega_R^2 +4\gamma\Gamma^2) + \omega_0\left(\Omega_R^2\Gamma(\Gamma + \gamma) -2\gamma(2\gamma^3 + 5\kappa\gamma^2-4\delta^2\kappa+4\gamma\kappa^2 + \kappa^3)\right)}
{\left(-\Omega_R^4+4\gamma(4\delta^2+\Gamma^2)\right)\left(\Omega_R^2\Gamma^2+\gamma\kappa(4\delta^2+\Gamma^2)\right)},
\end{equation}
\end{widetext}
where we have defined the detuning $\delta = \omega_{sp}-\omega_0$. The case of zero detuning $\delta \to 0$ reduces the above expression to Eq. (2) in the main text.

\end{document}